\documentclass[sigconf]{acmart}

\renewcommand\footnotetextcopyrightpermission[1]{} 
\usepackage[utf8]{inputenc}
\usepackage[T1]{fontenc}
\usepackage{booktabs} 
\usepackage{caption}
\usepackage{multirow}
\usepackage{graphicx}
\usepackage{enumitem}

\setcopyright{rightsretained}

\acmConference[Computation+Journalism]{Computation+Journalism Symposium}{October 2017}{Illinois, USA} 
\acmYear{2017}
\copyrightyear{2017}

\begin{document}
\title{A Large-scale Study of Social Media Sources in News Articles}

\author{Md Main Uddin Rony$^1$, Mohammad Yousuf$^2$, Naeemul Hassan$^1$}

\affiliation{$^1$Department of Computer Science and Engineering, The University of Mississippi}
\affiliation{$^2$Gaylord College of Journalism and Mass Communication, The University of Oklahoma}

\renewcommand{\shorttitle}{Use of Social Media Content as Source}

\renewcommand{\shortauthors}{Md Main Uddin et al.}
\setlist[itemize]{leftmargin=*}

\begin{abstract}
In this study, we closely look at the use of social media contents as source or reference in the U.S. news media. Specifically, we examine about $60$ thousand news articles published within the 5 years period of $2013-2017$ by $153$ U.S. media outlets and analyze use of social media content as source compared to other sources. We designed a social media source extraction algorithm and investigated the extent and nature of social media source usage across different news topics. Our results show that uses of social media content in news almost doubled in five years. Unreliable media outlets rely on social media more than the mainstream media. Both mainstream and unreliable sites prefer \textit{Twitter} to \textit{Facebook} as a source of information.
\end{abstract}

\maketitle

\newcommand{\reminder}[1]{\textbf{\textit{Reminder: }\color{blue}{#1}}}
\newcommand{\updatethis}[1]{\textbf{\color{red}{#1}}}


\newcommand{\dataset}[1]{%
    \IfEqCase{#1}{%
        {1}{Headlines}%
        {2}{Media Corpus}%
    }[\PackageError{tree}{Undefined option to tree: #1}{}]%
}%

\newcommand*{\rom}[1]{\expandafter\@slowromancap\romannumeral #1@}

\newcommand{\googlewv}{Google\_word2vec}
\newcommand{\clkbtwv}{Skip-Gram$_{sw}$~}
\newcommand{\glove}{GLOVE}

\newcommand{\subsubsubsection}[1]{\noindent\textbf{#1}:~}

\section{Introduction}
\label{motivation}

Uses of social media content (e.g., Tweets and Facebook posts) in news stories have become a common practice in newsrooms across the world \cite{broersma2013twitter,hladik2017powers,paulussen2014social}). Journalists quote and paraphrase contents regularly from social media pages. For instance, an article from the National Broadcasting Company (NBC) news~\footnote{https://www.nbcnews.com/politics/donald-trump/trump-slams-release-secretly-recorded-cohen-conversation-so-sad-n894401} says- \textit{``What kind of a lawyer would tape a client? So sad! Is this a first, never heard of it before?'' Trump tweeted.} This article has used a social media content (in this case, a Tweet) as a source. According to \cite{broersma2013twitter}, social media contents are being used as source because it is \textit{``convenient, cheap and effective''}. Some researchers have investigated the extent to which mainstream news media in some European countries used Facebook, Twitter and YouTube contents in news (~\cite{broersma2013twitter,hladik2017powers,paulussen2014social}). However, such a study on U.S. news media is absent. Also the previous studies were performed over small sample size which limits the scope of the findings. The purpose of this large-scale study is to examine the extent to which mainstream U.S. based news media use social media content (Facebook and Twitter) as sources of information. We also examine similar practices on many online news portals that are popular but considered by many~\cite{wikifakelist,melissa2016m,info2016list} as unreliable. We compare the social media source usage with respect to traditional source usage. We further investigate the relation between social media source and news topic category. 

There is a set of challenges which we had to overcome to conduct this study. First, a large-scale dataset of news articles from mainstream and unreliable U.S. media outlets is not available. There are some datasets that cover only the headlines \cite{rony2017diving} or cover a limited number of media outlets \cite{nytimesdata}. However, these datasets are not adequate for this study as our objective is to examine patterns of social media content usage of a range of media categories over a reasonable period of time. For this reason, we carefully collected about $60$ thousand news articles which were published within the years $2013-2017$ from $153$ U.S. media outlets. The next challenge is to identify the used sources in the news articles. Due to the large-scale nature of the data, it is not feasible to extract source and quotes manually from the news articles. So, we depend on automatic source extraction.  
We design a rule-based classifier that automatically identifies whether a direct quote is sourced from Social Media (Facebook, Twitter) or not with $89.8\%$ precision. We further extend the classifier to identify paraphrased quotes (not direct) as well that achieves a precision of $94.34\%$. Using these classifiers, we process all the collected news articles and analyze underlying social media source usage patterns within mainstream and unreliable media over the time. Our analysis shows that the practice of sourcing from Facebook and Twitter has doubled within the five years period. In a nutshell, we make the following contributions in this paper- \textbf{i)} we prepared a large dataset of news articles published by mainstream and unreliable U.S. media; there is plan to share this dataset with the research community upon acceptance of the manuscript, \textbf{ii)} we developed an automated social media source classifier and evaluated its performance, \textbf{iii)} using the dataset and the classifier, we analyze the social media source usage patterns in U.S. media. According to our knowledge, no other researchers have explored this before.

This interdisciplinary study is a substantial addition to the literature on the journalistic use of social media as it relates to sourcing practices. Despite the importance of research on this topic articulated in scholarly and professional discussions, no study examined the practices in U.S. media. The current study seeks to fill that gap. By examining this new sourcing practice, the study provides an in-depth look at how social media contents are shaping public discourses in the United States.
\vspace{-2mm}
\section{Related Work}
\label{relatedwork}

\textbf{Influences on Sourcing Routine}:
A number of endogenous and exogenous factors, described by ~\cite{shoemaker2013mediating} as a hierarchy of influences, determine the process of news production. Several scholars used the \textit{Hierarchy of Influences} model to explain sourcing practices in newsrooms ~\cite{kruikemeier2018news,turcotte2017s,yamamoto2017us}. Some key factors behind source selection include personal relationship, relevance, accessibility or willingness of a source to talk to a reporter, and credibility \cite{gans1979deciding, reich2011source}.
~\cite{gans1979deciding} suggested that a combination of economic and authoritative considerations aimed at producing quality news coverage with limited resources determined the source selection processes in traditional newsrooms such as newspapers, magazines, and network news. The economic consideration refers to efficiency or optimal use of available resources while the
authoritative consideration refers to the perceived degree of authority attributed to sources ~\cite{paulussen2014social}. These considerations often lead journalists to choose sources from a
small pool of known sources--mostly government officials and the powerful elites--who had already established their credibility and relevance \cite{kruikemeier2018news,lasorsa1990news,reich2011source}. 
Social hierarchy appeared in the literature as an undisputed predictor of news sourcing practices. All major theoretical frameworks used in the studies on news sourcing patterns--~\cite{shoemaker2013mediating} `hierarchy of influences', ~\cite{gans1979deciding} `hierarchies of nation and society', and ~\cite{becker1967whose} `hierarchy of credibility' --point to the same conclusion: Journalists rely more on institutional sources than ordinary people ~\cite{thurman2008forums,reich2011source}. Though credibility was apparently the most important factor behind source selection, resource constraints of news organizations and easier access to media had played a key role in establishing elite dominance in news ~\cite{hermida2014sourcing,gans1979deciding,reich2011source}.

\textbf{Integration of Social Media in Sourcing}: The Internet and new technologies--particularly social media--offer news reporters easy access to a large pool of diverse sources who would otherwise be hard to approach ~\cite{broersma2013twitter,hermida2010twittering}. Various studies show that news reporters are increasingly integrating social media content in news ~\cite{broersma2013twitter,chadwick2017hybrid,messner2008source,parmelee2011politics,paulussen2014social,kruikemeier2018news}. Such content includes quotes and paraphrases from posts on social media such as blogs, Tweets, Facebook, and YouTube posts. A survey of British journalists examined how journalists in UK, Germany, Sweden, and Finland view and use social media ~\cite{gulyas2011perceptions}. The study found that nearly all journalists in the UK (97\%) use social media for work, but many journalists are skeptical of the reliability of social media content. News organizations use social media more for publishing and distributing content than for sourcing. Mainstream news media organizations (e.g., BBC, CNN, The New York Times, The Washington Post) have long been integrating social media content into news \cite{messner2008source}. \cite{paulussen2014social} did a study on Belgian newspapers while ~\cite{hladik2017powers} studied Czech newspapers and came up with evidence of the use of content from Facebook, Twitter, and YouTube. Though the use of social media content in news has become a trend, not all journalists see social media as an influential news source. \cite{hedman2013social,yamamoto2017us, gulyas2011perceptions} suggest that those who have longer professional experience and those who work for private newspapers consider social media as less influential or less newsworthy. On the other hand, younger editors coming from pluralistic society--especially the ones who work for publicly owned newspapers--see social media content as an influential source. ~\cite{lariscy2009examination} found that acceptability of social media as a source is low among business journalists compared to the general trend.
In sum, integration of social media content in news continued to grow despite some mild opposition from a section of journalists. Many reporters rely heavily on social media to find and verify information. Worsening financial situation of news organizations, as well as increasing user expectation for news on demand, will continue to force newsrooms to use social media as a source ~\cite{broersma2013twitter}. As the integration of social media content in news has become a \textit{fait accompli}, many scholars debated how this is reshaping the sourcing routine in newsrooms. ~\cite{van2017follow} identified two opposing views that dominated this debate. A section of the literature suggests that social media helped legacy news media diversify its sources and include more voices of ordinary citizens in news. It is, thus, replacing the existing power-to-people hierarchy with a bottom-up approach. Another group of scholars suggested that there was no change in traditional sourcing routine as elite sources, also known as experts ~\cite{freedman2010gender}, continued to dominate news media on the web.

\section{Research Questions}
\label{problem}
The existing literature covers various aspects of journalistic use of social media. But it lacks systematic research on how U.S. news media deploy social media content in news. This study seeks to address three major aspects of social media content use in news--frequency of use, processing of content, and relation to news topic--and asks the following research questions.

\begin{itemize}
\item \textbf{RQ1:} How often do mainstream and unreliable news websites use Facebook and Twitter
content in articles?
\item \textbf{RQ2:} To what extent do mainstream and unreliable news media process Facebook and Twitter content used in articles?
\item \textbf{RQ3:} Does the use of social media source vary for different news topics?
\end{itemize}

\section{Methodology}
\label{method}
\subsection{Data Collection}
We prepared a list of websites of mainstream and unreliable news media based
on circulation, rating and popularity among Internet users ~\cite{info2016list,schneider2016m,melissa2016m}. The list included websites of $68$ U.S. mainstream media ($25$ print news media, $43$ broadcast), and $85$ unreliable media. The unreliable media included websites described as conspiracy, clickbait, satire, and junk science by ~\cite{info2016list} and \cite{melissa2016m}. Further details about the mainstream and unreliable media selection process can be found in \cite{rony2017diving}. We followed the official Facebook pages of these media and using the Facebook Graph API, we collected up to $600$ posts per media per year. These posts were published on Facebook within January 1, 2013 and December 31, 2017. A total of $376,199$ Facebook posts were collected. A Facebook post may contain a photo or a video or a link to an external source. For each post, we collected the headline (title of a video or headline of an article), status type, link to the main article, and the status message. Of the total posts, only $135,294$ contained links to the corresponding news article. For each link, we used a publicly available Python package named Newspaper3k \footnote{https://newspaper.readthedocs.io} to collect the news article content. Some links lead to unavailable web page and some links were restricted due subscription limit. At the end, we had $29,656$ articles from mainstream media and $76,997$ from unreliable media.  We took a random sample of $29,700$ unreliable news and prepare a balanced dataset of $59,356$ news articles.

\subsection{Extraction of Social Media Source}
\begin{figure}[t!]
\vspace{-7mm}
\includegraphics[width=\linewidth]{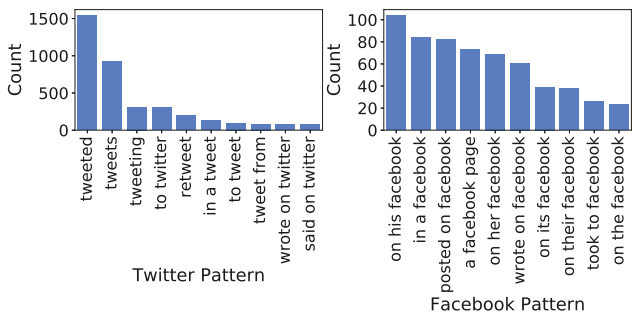}
\centering
\vspace{-8mm}
\caption{Most frequent Twitter and Facebook patterns}
\label{fig:fb_tw_most_frq_pat}
\end{figure}

Identifying social media content that is used as a source in a news article is a challenging task. Because, social media contents can be of different forms. For instance, the content can be a text, or image, or a video. In this study, we restricted ourselves to the text format only. Still, the linguistic variations of the way a Tweet or a Facebook post can be cited as a source posed a big challenge. For example, \emph{she tweeted}, \emph{the tweet read in part}, \emph{took to Twitter}- all these patterns can be used to cite a Twitter source. To identify these language variations, we examined a set of news articles manually and carefully curated a list of patterns. Specifically, we took a stratified random sample of $400$ news articles which contained any of the following keywords--\textit{facebook}, \textit{twitter}, \textit{post}, \textit{tweet}. Then, we manually inspected these articles and identified $212$ citation patterns ($134$ patterns for Facebook and $78$ patterns for Twitter) that were used to cite social media post as a source. Figure \ref{fig:fb_tw_most_frq_pat} shows some examples of these citation patterns. These patterns were used to automatically extract social media sources from the collected data.

\subsubsection{\textbf{Quotation, Paraphrase, and Embedding}}
\label{process}
We observe, a social media source is often directly quoted fully or partially (for instance, the NBC example in Section \ref{motivation}). Sometimes, a source is processed and paraphrased by the news reporter. For example, \emph{she tweeted that she was glad to have lost 6 pounds}~\footnote{https://www.smh.com.au/entertainment/celebrity/kim-kardashians-flu-weight-loss-tweet-should-have-been-celebrated-not-condemned-20170421-gvpw8l.html}. And sometimes, a source is directly embedded~\footnote{https://developer.twitter.com/en/docs/twitter-for-websites/embedded-tweets/overview.html} without quoting or paraphrasing. Embedding happens mostly in case of Twitter sources and they are easier to identify using a regular expression pattern. We use the above described patterns to identify a source usage and then categorize that into one of these three types- \textit{Quotation}, \textit{Paraphrase}, and \textit{Embedding}. Specifically, first, we segment an article into sentences using the NLTK~\footnote{https://www.nltk.org/} python package. Then, for each sentence, we check if it contains one of the $212$ patterns. If it matches the embedding regular expression, we categorize the source usage as an \emph{Embedding}. If it matches with other patterns, we examine if the sentence contains quotation signs (" " or `` '' or ` ' or other similar variants) or not. If it does, we categorize the source usage as a \emph{Quotation}. Otherwise, we categorize it as a \emph{Paraphrase}. We did not consider Facebook embedding as we found that Facebook posts were rarely embedded in web pages.

\begin{table}[t!]
\begin{tabular}{l@{\hspace{0.7in}}|r|r|r}
\toprule
Category		& Precision		& Recall		& F1		\\
\midrule
Quotation		& 89.80\%		& 73.33\%		& 86.21\% 	\\
Paraphrase		& 94.34\%		& 79.37\%		& 80.73\% 	\\
Embedding		& 100\%			& 100\%			& 100\% 	\\
\midrule
Macro-average	& 94.71\%		& 84.23\%		& 88.98\%	\\
Micro-average	& 97.85\%		& 92.62\%		& 95.16\%	\\
\bottomrule
\end{tabular}
\caption{Performance of Social Media Source Identification}
\label{tab:pr_sm_mention_iden}
\end{table}

\subsubsection{\textbf{Performance Evaluation}}
To evaluate the performance of the above explained social media source identification and categorization program, we randomly sampled 100 news articles where the program found at least one social media source and another 100 random news articles where the program didn't find any social media source. Then, we manually inspected the articles to identify the social media sources and their categories. Our manual analysis found that in total there were 393 social media source usages (60 \textit{Quotations}, 63 \textit{Paraphrases}, and 270 \textit{Embeddings}). Our program found in total 372 sources (49 \textit{Quotations}, 53 \textit{Paraphrases}, and 270 \textit{Embeddings}). Out of these 372 sources, 364 (44 \textit{Quotations}, 50 \textit{Paraphrases}, and 270 \textit{Embeddings}) were accurate and 8 were falsely marked as a social media source. Out of the 393 true media sources, the program could not identify 29 sources. Table ~\ref{tab:pr_sm_mention_iden} shows recall and precision of the program. As \emph{Embedding} follows a fixed pattern, the program could identify all the embeddings correctly. The reason behind lower recall for \emph{Quotation} and \emph{Paraphrase} is the use of uncommon linguistic patterns (e.g., ``put his two cents on Twitter'', ``137 characters of angry tweet'') by writers. To avoid over-fitting, we decided not to add these uncommon linguistic patterns in our list of patterns.

\section{Analysis}
\label{results}
Using the collected data and the developed source identification program, we answer the research questions posed in Section \ref{problem}.

\subsection{RQ1: Use of Social Media as Source}
RQ1 asks \textit{how often do mainstream and unreliable news websites use Facebook and Twitter content in articles?} We applied the social media source identification program on all the $59,356$ mainstream and unreliable articles. Table \ref{tab:sm_mention_by_media} shows results of this application. In total, we find that 5,430 articles (9.15\% of all data) contained at least one social media post as a source. Note that an article may contain both Facebook and Twitter source.
However, a major portion of the articles use Twitter (4,824) as a source compared to Facebook (701). The underlying reason could be that the Tweets are generally public whereas Facebook posts are not. Also, it is convenient to embed a Tweet whereas Facebook post embedding is rarely seen in news articles. Moreover, politicians and celebrities use Twitter more frequently than they do Facebook to engage with the people. We also find that the unreliable organizations use social media posts as a source more often than the mainstream media. Of the above mentioned 5,430 articles, 6.68\% belong to mainstream and 11.61\% (almost double) belong to unreliable. In total, there are 4,207 social media sources in 1,982 mainstream articles (2.12 source per article) and 12,436 sources in 3,448 unreliable articles (3.61 source per article). We observe that even though Twitter dominates Facebook in terms of source usage, the mainstream media use Facebook sources more often than unreliable media outlets. 10.29\% of all social media contents in mainstream articles are sourced from Facebook whereas only 2.85\% of social media sources in unreliable articles are from Facebook.

\begin{table*}[t]
\centering
\resizebox{\textwidth}{!}{
\begin{tabular}{l|r|r|r|r|r|r|r|r|r|r|r|r}
\hline
\multicolumn{1}{c|}{\textbf{Media Type}} & \multicolumn{1}{c|}{\textbf{\begin{tabular}[c]{@{}c@{}}Total \\ Articles\end{tabular}}} & \multicolumn{1}{c|}{\textbf{\begin{tabular}[c]{@{}c@{}}\#Articles Contain\\  SM Source\end{tabular}}} & \multicolumn{5}{c|}{\textbf{Twitter Source}}                                                                                                                                                                                                                       & \multicolumn{4}{c|}{\textbf{Facebook Source}}                                                                                                                                                                             & \multicolumn{1}{c}{\textbf{\begin{tabular}[c]{@{}c@{}}Total \\ Source\end{tabular}}} \\ \hline
\multicolumn{1}{c|}{}                    & \multicolumn{1}{c|}{}                                                                   & \multicolumn{1}{c|}{}                                                                                  & \multicolumn{1}{c|}{\textbf{\# of Articles}} & \multicolumn{1}{c|}{\textbf{Quotation}} & \multicolumn{1}{c|}{\textbf{Paraphrase}} & \multicolumn{1}{c|}{\textbf{Embedding}} & \multicolumn{1}{c|}{\textbf{\begin{tabular}[c]{@{}c@{}}Total\end{tabular}}} & \multicolumn{1}{c|}{\textbf{\# of Articles}} & \multicolumn{1}{c|}{\textbf{Quotation}} & \multicolumn{1}{c|}{\textbf{Paraphrase}} & \multicolumn{1}{c|}{\textbf{\begin{tabular}[c]{@{}c@{}}Total\end{tabular}}} & \multicolumn{1}{l}{}                                                                 \\ \cline{4-12}
Mainstream                               & 29656                                                                                   & 1982 (6.68\%)                                                                                          & 1654                                         & 1065 (28.22\%)                          & 866 (22.95\%)                            & 1843 (48.83\%)                          & 3774 (89.71\%)                                                                        & 377                                          & 228 (52.66\%)                           & 205 (47.34\%)                            & 433 (10.29\%)                                                                          & 4207                                                                                 \\
Unreliable                               & 29700                                                                                   & 3448 (11.61\%)                                                                                         & 3170                                         & 1137 (9.41\%)                           & 1130 (9.35\%)                            & 9814 (81.24\%)                          & 12081 (97.15\%)                                                                       & 324                                          & 178 (50.14\%)                           & 177 (49.86\%)                            & 355 (2.85\%)                                                                           & 12436                                                                                \\ \hline
\textbf{Total}                           & 59356                                                                                   & 5430 (9.15\%)                                                                                          & 4824 (88.84\%)                               & 2202                                    & 1996                                     & 11657                                   & 15855                                                                                 & 701                                          & 406                                     & 382                                      & 788                                                                                    & 16643                                                                                \\ \hline
\end{tabular}
}
\caption{Social media (Twitter and Facebook) content usage as a source by mainstream and unreliable media}
\label{tab:sm_mention_by_media}
\end{table*}

\begin{table*}[t!]
\centering
\vspace{-5mm}
\resizebox{\linewidth}{!}{
\begin{tabular}{l|r|r|r|r|r|r|r|r|r}
\hline
\multicolumn{1}{c|}{\textbf{Topic}} & \multicolumn{1}{c|}{\textbf{\# Articles with Social Media Source}} & \multicolumn{4}{c|}{\textbf{Mainstream}}                                                                                                                                 & \multicolumn{4}{c}{\textbf{Unreliable}}                                                                                                                                \\ \hline
                                    & \multicolumn{1}{l|}{}                                       & \multicolumn{1}{c|}{\textbf{\# Articles}} & \multicolumn{1}{c|}{\textbf{Quotation}} & \multicolumn{1}{c|}{\textbf{Paraphrase}} & \multicolumn{1}{c|}{\textbf{Embedding}} & \multicolumn{1}{c|}{\textbf{\# Articles}} & \multicolumn{1}{c|}{\textbf{Quotation}} & \multicolumn{1}{c|}{\textbf{Paraphrase}} & \multicolumn{1}{c}{\textbf{Embedding}} \\ \hline
Politics                            & 2080                                                        & 369                                       & 377 (47.72\%)                           & 238 (30.13\%)                            & 175 (22.15\%)                           & 1711                                      & 665 (9.77\%)                            & 648 (9.52\%)                             & 5495 (80.71\%)                         \\ \hline
Arts \& Entertainment               & 798                                                         & 491                                       & 303 (26.84\%)                           & 241 (21.35\%)                            & 585 (51.82\%)                           & 307                                       & 197 (14.71\%)                           & 116 (8.66\%)                             & 1026 (76.62\%)                         \\ \hline
Sensitive Subject                   & 784                                                         & 300                                       & 187 (32.86\%)                           & 165 (29\%)                               & 217 (38.14\%)                           & 484                                       & 217 (13.79\%)                           & 172 (10.93\%)                            & 1185 (75.28\%)                         \\ \hline
Law \& Government                   & 340                                                         & 112                                       & 69 (34.85\%)                            & 73 (36.87\%)                             & 56 (28.28\%)                            & 228                                       & 75 (11.28\%)                            & 74 (11.13\%)                             & 516 (77.59\%)                          \\ \hline
Sports                              & 283                                                         & 213                                       & 97 (17.05\%)                            & 380 (66.78\%)                            & 92 (16.17\%)                            & 70                                        & 30 (17.05\%)                            & 28 (15.91\%)                             & 118 (67.04\%)                          \\ \hline
People \& Society                   & 239                                                         & 68                                        & 22 (18.18\%)                            & 38 (31.40\%)                             & 61 (50.41\%)                            & 171                                       & 64 (11.79\%)                            & 66 (12.15\%)                             & 413 (76.06\%)                          \\ \hline
Health                              & 76                                                          & 25                                        & 11 (39.29\%)                            & 9 (32.14\%)                              & 8 (28.57\%)                             & 51                                        & 21 (18.58\%)                            & 26 (23.01\%)                             & 66 (58.41\%)                           \\ \hline
\end{tabular}%
}
\caption{Processing of social media sources in different news topics}
\label{tab:topic-smmention-distribution}
\end{table*}

\begin{figure}
\vspace{-7mm}
\includegraphics[width=\linewidth]{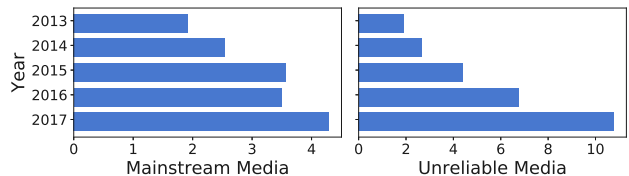}
\centering
\vspace{-7mm}
\caption{Social media source usage is increasing over time}
\label{fig:use_sm_content}
\end{figure}

We further examine how the social media source usage has evolved over the time. We categorized the articles per year. The distribution of articles over the years is as follows- '2013': 7,176, '2014': 10,725, '2015': 14,585, '2016': 12,694, '2017': 14,176. We observe that the practice of citing social media content is increasing over time. For example, in 2013, about 3.85\% articles (276 out of 7,176) used social media as a source whereas in 2017, the percentage was about 15.05\% (2,134 out of 14,176 articles). For each year, figure ~\ref{fig:use_sm_content} shows the percentage of articles from mainstream and unreliable media that use social media (Facebook/Twitter) as a source. The practice is increasing in both categories. 

\begin{table}[t]
\vspace{-5mm}
\centering 
\begin{tabular}{l|r|r|r}
\hline
\multicolumn{1}{c|}{\textbf{Media Type}} & \multicolumn{1}{c|}{\textbf{\begin{tabular}[c]{@{}c@{}}\# Direct Quote \\ (Avg. Per Article)\end{tabular}}} & \multicolumn{1}{c|}{\textbf{\begin{tabular}[c]{@{}c@{}}\# Social Media \\ Source\end{tabular}}} & \multicolumn{1}{c}{\textbf{\begin{tabular}[c]{@{}c@{}}Ratio\end{tabular}}} \\ \hline
Mainstream                               & 201924 (6.81)                                                                                               & 4207                                                                                   & 1:48                                                                                                       \\ \hline
Unreliable                               & 185182 (6.23)                                                                                               & 12436                                                                                  & 1:14.89                                                                                                    \\ \hline
\end{tabular}
\caption{Social media source vs. all direct quotations}
\label{tab:sm_direct_quote}
\end{table}

We also study the extent of social media source usage with respect to all kinds of sources including regular, non-social media based sources (e.g., interview, book, press release). For instance, \emph{``I'm just going to pay my respects,'' Trump told Fox News on Monday night}~\footnote{https://www.reuters.com/article/us-pennsylvania-shooting/trump-to-visit-pittsburgh-amid-funerals-calls-for-him-to-stay-away-idUSKCN1N418P})-- is an example of a direct quote from a source which is not from social media. Automatically extracting all kinds of sources is a very challenging task as a source can be directly quoted or paraphrased using many linguistic variations. In this study, we restricted the comparison among direct quotes only which are relatively easier to extract automatically. Details of our direct quote extraction method can be found in \cite{muzny2017twostage,manning-EtAl:2014:P14-5}. Table \ref{tab:sm_direct_quote} shows the comparison between use of direct quotations and social media source. On average, a mainstream article uses more direct quotations (6.81 quotes per article) compared to an unreliable article (6.23 quotes per article). However, an average unreliable article use one social media source for every 14.89 direct quotes where an average mainstream article uses significantly lower number of social media sources (one for every 48 direct quotes).


\subsection{RQ2: Processing of Social Media Content}
RQ2 asks \textit{to what extent do mainstream and unreliable news media process Facebook and Twitter content used in articles?} We wanted to understand whether the media outlets embeds a source (\emph{Embedding}), or directly quote a source (\emph{Quotation}), or paraphrases a source (\emph{Paraphrase}) . Using the program described in Section \ref{process}, we examine how media processes a source. Table \ref{tab:sm_mention_by_media} summarizes the results. We observe that in case of Twitter, both media tends to process the sources more as \emph{Embeddings} rather than \emph{Quotation} or \emph{Paraphrase}. The unreliable media uses \emph{Quotation} and \emph{Paraphrase} almost equally whereas mainstream media uses more \emph{Quotation} than \emph{Paraphrase}.
In case of Facebook sources, the distribution of \emph{Quotation} versus \emph{Paraphrase} is more balanced. We also infer that irrespective of social platforms (Twitter, Facebook), mainstream media uses more social media source as \emph{Quotations} compared to unreliable media, though the difference is significant for Twitter.


\begin{table}[t!]
\vspace{-5mm}
\centering
\resizebox{\linewidth}{!}{
\begin{tabular}{l|l|r|r}
\hline
\multicolumn{1}{c|}{\textbf{Media Type}} & \multicolumn{1}{c|}{\textbf{Topic}} & \multicolumn{1}{c|}{\textbf{\begin{tabular}[c]{@{}c@{}}\# Articles\end{tabular}}} & \multicolumn{1}{c}{\textbf{\begin{tabular}[c]{@{}c@{}}\# Articles with \\ Social Media Source\end{tabular}}} \\ \hline
\multirow{5}{*}{Mainstream}              & Arts \& Entertainment               & 5943                                                                                    & 491 (8.26\%)                                                                                              \\
                                         & Sensitive Subjects                  & 3391                                                                                    & 300 (8.85\%)                                                                                              \\
                                         & Law \& Government                   & 2793                                                                                    & 112 (4.01\%)                                                                                              \\
                                         & Sports                              & 2592                                                                                    & 213 (8.22\%)                                                                                              \\
                                         & Politics                            & 2389                                                                                    & 369 (15.45\%)                                                                                             \\ \hline
\multirow{5}{*}{Unreliable}              & Politics                            & 7104                                                                                    & 1711 (24.09\%)                                                                                            \\
                                         & Sensitive Subjects                  & 3790                                                                                    & 484 (12.77\%)                                                                                             \\
                                         & People \& Society                   & 2889                                                                                    & 171 (5.92\%)                                                                                              \\
                                         & Law \& Government                   & 2835                                                                                    & 228 (8.04\%)                                                                                              \\
                                         & Health                              & 2546                                                                                    & 51 (2\%)                                                                                                  \\ \hline
\end{tabular}
}
\caption{Extent of social media source usage in the top-5 news topics for each media}
\label{tab:topic-article-distribution}
\end{table}


\subsection{RQ3: Relation With News Topic}
RQ3 asks \emph{does the use of social media source vary for different news topics?} To answer this question, we categorize each of the $59,356$ articles into $27$ topics using Google Cloud Natural Language API~\footnote{https://cloud.google.com/natural-language/}. A complete list of the topics can be found here \url{https://cloud.google.com/natural-language/docs/categories}. Table \ref{tab:topic-article-distribution} shows the top-5 topics for each media category. These topics cover 58\% and 64\% of all the mainstream and unreliable articles, respectively. Both media use social media source in \emph{Politics} related news more often than in other news topics. Also, in all the three common topics (\emph{Sensitive Subjects, Law \& Government, and Politics}), unreliable media uses more social media source compared to mainstream media. 
We further investigate the processing of social media source in the top-5 topics. Table \ref{tab:topic-smmention-distribution} shows the distribution of social media source processing categories (\emph{Quotation}, \emph{Paraphrase}, \emph{Embedding}) for these seven (union of top-5 topics in each media category) topics-- Politics, Arts \& Entertainment, Sensitive Subject, Law \& Government, Sports, People \& Society, and Health. In all these seven categories, unreliable media processes social media sources as \emph{Embeddings} significantly more compared to mainstream media. On the other hand, mainstream media uses \emph{Quotation} more often compared to unreliable media in six of these seven categories.

\section{Discussion}
\label{discussion}

The main objective of this study is to provide a glimpse of how deeply social media content has been rooted in news. It sought to understand the extent to which online news media uses Facebook and Twitter as a source of information. It also compared differences between mainstream news media and unreliable media as they use social media as a news gathering tool. The study has been longitudinal in nature since it provides an overview of five years. Following is the discussion of some major findings.

First, the data identified several patterns that support previous studies suggesting Twitter, among other social networking sites, is the preferred source of information to online news and informational content creators ~\cite{broersma2013twitter,paulussen2014social}. For instance, the findings suggest that both mainstream and unreliable news websites generally prefer Twitter to Facebook as a news source. Second, we find that the number of social media sources in news is increasing by a large number every year. This number has almost doubled in five years. This finding confirms previous research suggesting that uses of social media in news were increasing ~\cite{broersma2013twitter,chadwick2017hybrid,kruikemeier2018news,messner2008source,parmelee2011politics,paulussen2014social}. Third, the results show that unreliable websites are more dependent on social media than the mainstream media, which speaks of their weak organizational structure and lack of resources to produce quality news content ~\cite{atton2005sourcing}. This also supports the assumption that they rely on free and biased content to fill pages and shore up their agenda. In general, the study has several contributions to the journalism literature. First, it gives an overview of social media content uses in U.S. Second, the study examined five years of data showing a steady increase in citing social media sources over time. Previous research only assumed this, but the current study conducted a systematic study on a large data set. Third, this study also examined practices of unreliable media websites that are rarely studied.

However, this study has several limitations. For instance, we only considered Facebook and Twitter as social media platforms where there are many other popular sites as well such as Instagram, YouTube, Tumblr, etc. However, we believe Facebook and Twitter are the most prominent social networks used nowadays. Also, our source and quotation identification methods are simple, yet accurate, as explained in the method section. These could be further improved by applying machine learning and natural language processing techniques. In future, we want to overcome these limitations and at the same time explore other possible directions. 


\bibliographystyle{ACM-Reference-Format}
\bibliography{reference} 


\begin{thebibliography}{00}


\ifx \showCODEN    \undefined \def \showCODEN     #1{\unskip}     \fi
\ifx \showDOI      \undefined \def \showDOI       #1{#1}\fi
\ifx \showISBNx    \undefined \def \showISBNx     #1{\unskip}     \fi
\ifx \showISBNxiii \undefined \def \showISBNxiii  #1{\unskip}     \fi
\ifx \showISSN     \undefined \def \showISSN      #1{\unskip}     \fi
\ifx \showLCCN     \undefined \def \showLCCN      #1{\unskip}     \fi
\ifx \shownote     \undefined \def \shownote      #1{#1}          \fi
\ifx \showarticletitle \undefined \def \showarticletitle #1{#1}   \fi
\ifx \showURL      \undefined \def \showURL       {\relax}        \fi
\providecommand\bibfield[2]{#2}
\providecommand\bibinfo[2]{#2}
\providecommand\natexlab[1]{#1}
\providecommand\showeprint[2][]{arXiv:#2}

\bibitem[\protect\citeauthoryear{Atton and Wickenden}{Atton and
  Wickenden}{2005}]%
        {atton2005sourcing}
\bibfield{author}{\bibinfo{person}{Chris Atton} {and} \bibinfo{person}{Emma
  Wickenden}.} \bibinfo{year}{2005}\natexlab{}.
\newblock \showarticletitle{Sourcing routines and representation in alternative
  journalism: A case study approach}.
\newblock \bibinfo{journal}{{\em Journalism Studies\/}} \bibinfo{volume}{6},
  \bibinfo{number}{3} (\bibinfo{year}{2005}), \bibinfo{pages}{347--359}.
\newblock


\bibitem[\protect\citeauthoryear{Becker}{Becker}{1967}]%
        {becker1967whose}
\bibfield{author}{\bibinfo{person}{Howard~S Becker}.}
  \bibinfo{year}{1967}\natexlab{}.
\newblock \showarticletitle{Whose side are we on?}
\newblock \bibinfo{journal}{{\em Social problems\/}} \bibinfo{volume}{14},
  \bibinfo{number}{3} (\bibinfo{year}{1967}), \bibinfo{pages}{239--247}.
\newblock


\bibitem[\protect\citeauthoryear{Broersma and Graham}{Broersma and
  Graham}{2013}]%
        {broersma2013twitter}
\bibfield{author}{\bibinfo{person}{Marcel Broersma} {and} \bibinfo{person}{Todd
  Graham}.} \bibinfo{year}{2013}\natexlab{}.
\newblock \showarticletitle{Twitter as a news source: How Dutch and British
  newspapers used tweets in their news coverage, 2007--2011}.
\newblock \bibinfo{journal}{{\em Journalism practice\/}} \bibinfo{volume}{7},
  \bibinfo{number}{4} (\bibinfo{year}{2013}), \bibinfo{pages}{446--464}.
\newblock


\bibitem[\protect\citeauthoryear{Chadwick}{Chadwick}{2017}]%
        {chadwick2017hybrid}
\bibfield{author}{\bibinfo{person}{Andrew Chadwick}.}
  \bibinfo{year}{2017}\natexlab{}.
\newblock \bibinfo{booktitle}{{\em The hybrid media system: Politics and
  power}}.
\newblock \bibinfo{publisher}{Oxford University Press}.
\newblock


\bibitem[\protect\citeauthoryear{Freedman, Fico, and Durisin}{Freedman
  et~al\mbox{.}}{2010}]%
        {freedman2010gender}
\bibfield{author}{\bibinfo{person}{Eric Freedman}, \bibinfo{person}{Frederick
  Fico}, {and} \bibinfo{person}{Megan Durisin}.}
  \bibinfo{year}{2010}\natexlab{}.
\newblock \showarticletitle{Gender diversity absent in expert sources for
  elections}.
\newblock \bibinfo{journal}{{\em Newspaper Research Journal\/}}
  \bibinfo{volume}{31}, \bibinfo{number}{2} (\bibinfo{year}{2010}),
  \bibinfo{pages}{20--33}.
\newblock


\bibitem[\protect\citeauthoryear{Gans}{Gans}{1979}]%
        {gans1979deciding}
\bibfield{author}{\bibinfo{person}{Herbert Gans}.}
  \bibinfo{year}{1979}\natexlab{}.
\newblock \showarticletitle{Deciding What's News (New York}.
\newblock \bibinfo{journal}{{\em Pantheon\/}}  \bibinfo{volume}{241}
  (\bibinfo{year}{1979}).
\newblock


\bibitem[\protect\citeauthoryear{Guly{\'a}s}{Guly{\'a}s}{2011}]%
        {gulyas2011perceptions}
\bibfield{author}{\bibinfo{person}{A Guly{\'a}s}.}
  \bibinfo{year}{2011}\natexlab{}.
\newblock \bibinfo{title}{Perceptions and Use of Social Media among Journalists
  in the UK}.
\newblock   (\bibinfo{year}{2011}).
\newblock


\bibitem[\protect\citeauthoryear{Hedman and Djerf-Pierre}{Hedman and
  Djerf-Pierre}{2013}]%
        {hedman2013social}
\bibfield{author}{\bibinfo{person}{Ulrika Hedman} {and} \bibinfo{person}{Monika
  Djerf-Pierre}.} \bibinfo{year}{2013}\natexlab{}.
\newblock \showarticletitle{The social journalist: Embracing the social media
  life or creating a new digital divide?}
\newblock \bibinfo{journal}{{\em Digital Journalism\/}} \bibinfo{volume}{1},
  \bibinfo{number}{3} (\bibinfo{year}{2013}), \bibinfo{pages}{368--385}.
\newblock


\bibitem[\protect\citeauthoryear{Hermida}{Hermida}{2010}]%
        {hermida2010twittering}
\bibfield{author}{\bibinfo{person}{Alfred Hermida}.}
  \bibinfo{year}{2010}\natexlab{}.
\newblock \showarticletitle{Twittering the news: The emergence of ambient
  journalism}.
\newblock \bibinfo{journal}{{\em Journalism practice\/}} \bibinfo{volume}{4},
  \bibinfo{number}{3} (\bibinfo{year}{2010}), \bibinfo{pages}{297--308}.
\newblock


\bibitem[\protect\citeauthoryear{Hermida, Lewis, and Zamith}{Hermida
  et~al\mbox{.}}{2014}]%
        {hermida2014sourcing}
\bibfield{author}{\bibinfo{person}{Alfred Hermida}, \bibinfo{person}{Seth~C
  Lewis}, {and} \bibinfo{person}{Rodrigo Zamith}.}
  \bibinfo{year}{2014}\natexlab{}.
\newblock \showarticletitle{Sourcing the Arab Spring: A case study of Andy
  Carvin's sources on Twitter during the Tunisian and Egyptian revolutions}.
\newblock \bibinfo{journal}{{\em Journal of Computer-Mediated Communication\/}}
  \bibinfo{volume}{19}, \bibinfo{number}{3} (\bibinfo{year}{2014}),
  \bibinfo{pages}{479--499}.
\newblock


\bibitem[\protect\citeauthoryear{Hlad{\'\i}k and
  {\v{S}}t{\v{e}}tka}{Hlad{\'\i}k and {\v{S}}t{\v{e}}tka}{2017}]%
        {hladik2017powers}
\bibfield{author}{\bibinfo{person}{Radim Hlad{\'\i}k} {and}
  \bibinfo{person}{V{\'a}clav {\v{S}}t{\v{e}}tka}.}
  \bibinfo{year}{2017}\natexlab{}.
\newblock \showarticletitle{The powers that tweet: Social media as news sources
  in the Czech Republic}.
\newblock \bibinfo{journal}{{\em Journalism Studies\/}} \bibinfo{volume}{18},
  \bibinfo{number}{2} (\bibinfo{year}{2017}), \bibinfo{pages}{154--174}.
\newblock


\bibitem[\protect\citeauthoryear{informationisbeautiful.net}{informationisbeautiful.net}{2016}]%
        {info2016list}
\bibfield{author}{\bibinfo{person}{informationisbeautiful.net}.}
  \bibinfo{year}{2016}\natexlab{}.
\newblock \bibinfo{title}{Unreliable/Fake News Sites \& Sources}.
\newblock
  \bibinfo{howpublished}{\url{https://docs.google.com/spreadsheets/d/1xDDmbr54qzzG8wUrRdxQl_C1dixJSIYqQUaXVZBqsJs}}.
    (\bibinfo{year}{2016}).
\newblock


\bibitem[\protect\citeauthoryear{Kruikemeier and Lecheler}{Kruikemeier and
  Lecheler}{2018}]%
        {kruikemeier2018news}
\bibfield{author}{\bibinfo{person}{Sanne Kruikemeier} {and}
  \bibinfo{person}{Sophie Lecheler}.} \bibinfo{year}{2018}\natexlab{}.
\newblock \showarticletitle{News consumer perceptions of new journalistic
  sourcing techniques}.
\newblock \bibinfo{journal}{{\em Journalism Studies\/}} \bibinfo{volume}{19},
  \bibinfo{number}{5} (\bibinfo{year}{2018}), \bibinfo{pages}{632--649}.
\newblock


\bibitem[\protect\citeauthoryear{Lariscy, Avery, Sweetser, and Howes}{Lariscy
  et~al\mbox{.}}{2009}]%
        {lariscy2009examination}
\bibfield{author}{\bibinfo{person}{Ruthann~Weaver Lariscy},
  \bibinfo{person}{Elizabeth~Johnson Avery}, \bibinfo{person}{Kaye~D Sweetser},
  {and} \bibinfo{person}{Pauline Howes}.} \bibinfo{year}{2009}\natexlab{}.
\newblock \showarticletitle{An examination of the role of online social media
  in journalists’ source mix}.
\newblock \bibinfo{journal}{{\em Public relations review\/}}
  \bibinfo{volume}{35}, \bibinfo{number}{3} (\bibinfo{year}{2009}),
  \bibinfo{pages}{314--316}.
\newblock


\bibitem[\protect\citeauthoryear{Lasorsa and Reese}{Lasorsa and Reese}{1990}]%
        {lasorsa1990news}
\bibfield{author}{\bibinfo{person}{Dominic~L Lasorsa} {and}
  \bibinfo{person}{Stephen~D Reese}.} \bibinfo{year}{1990}\natexlab{}.
\newblock \showarticletitle{News source use in the crash of 1987: A study of
  four national media}.
\newblock \bibinfo{journal}{{\em Journalism Quarterly\/}} \bibinfo{volume}{67},
  \bibinfo{number}{1} (\bibinfo{year}{1990}), \bibinfo{pages}{60--71}.
\newblock


\bibitem[\protect\citeauthoryear{Manning, Surdeanu, Bauer, Finkel, Bethard, and
  McClosky}{Manning et~al\mbox{.}}{2014}]%
        {manning-EtAl:2014:P14-5}
\bibfield{author}{\bibinfo{person}{Christopher~D. Manning},
  \bibinfo{person}{Mihai Surdeanu}, \bibinfo{person}{John Bauer},
  \bibinfo{person}{Jenny Finkel}, \bibinfo{person}{Steven~J. Bethard}, {and}
  \bibinfo{person}{David McClosky}.} \bibinfo{year}{2014}\natexlab{}.
\newblock \showarticletitle{The {Stanford} {CoreNLP} Natural Language
  Processing Toolkit}. In \bibinfo{booktitle}{{\em Association for
  Computational Linguistics (ACL) System Demonstrations}}.
  \bibinfo{pages}{55--60}.
\newblock
\showURL{%
\url{http://www.aclweb.org/anthology/P/P14/P14-5010}}


\bibitem[\protect\citeauthoryear{Messner and Distaso}{Messner and
  Distaso}{2008}]%
        {messner2008source}
\bibfield{author}{\bibinfo{person}{Marcus Messner} {and}
  \bibinfo{person}{Marcia~Watson Distaso}.} \bibinfo{year}{2008}\natexlab{}.
\newblock \showarticletitle{The source cycle: How traditional media and weblogs
  use each other as sources}.
\newblock \bibinfo{journal}{{\em Journalism Studies\/}} \bibinfo{volume}{9},
  \bibinfo{number}{3} (\bibinfo{year}{2008}), \bibinfo{pages}{447--463}.
\newblock


\bibitem[\protect\citeauthoryear{Muzny, Fang, Chang, and Jurafsky}{Muzny
  et~al\mbox{.}}{2017}]%
        {muzny2017twostage}
\bibfield{author}{\bibinfo{person}{Grace Muzny}, \bibinfo{person}{Michael
  Fang}, \bibinfo{person}{Angel~X. Chang}, {and} \bibinfo{person}{Dan
  Jurafsky}.} \bibinfo{year}{2017}\natexlab{}.
\newblock \showarticletitle{A Two-stage Sieve Approach for Quote Attribution}.
  In \bibinfo{booktitle}{{\em Proceedings of the European Chapter of the
  Association for Computational Linguistics (EACL)}}.
\newblock
\showURL{%
\url{https://nlp.stanford.edu/pubs/muzny2017twostage.pdf}}


\bibitem[\protect\citeauthoryear{Parmelee and Bichard}{Parmelee and
  Bichard}{2011}]%
        {parmelee2011politics}
\bibfield{author}{\bibinfo{person}{John~H Parmelee} {and}
  \bibinfo{person}{Shannon~L Bichard}.} \bibinfo{year}{2011}\natexlab{}.
\newblock \bibinfo{booktitle}{{\em Politics and the Twitter revolution: How
  tweets influence the relationship between political leaders and the public}}.
\newblock \bibinfo{publisher}{Lexington Books}.
\newblock


\bibitem[\protect\citeauthoryear{Paulussen and Harder}{Paulussen and
  Harder}{2014}]%
        {paulussen2014social}
\bibfield{author}{\bibinfo{person}{Steve Paulussen} {and}
  \bibinfo{person}{Raymond~A Harder}.} \bibinfo{year}{2014}\natexlab{}.
\newblock \showarticletitle{Social media references in newspapers: Facebook,
  Twitter and YouTube as sources in newspaper journalism}.
\newblock \bibinfo{journal}{{\em Journalism Practice\/}} \bibinfo{volume}{8},
  \bibinfo{number}{5} (\bibinfo{year}{2014}), \bibinfo{pages}{542--551}.
\newblock


\bibitem[\protect\citeauthoryear{Reich}{Reich}{2011}]%
        {reich2011source}
\bibfield{author}{\bibinfo{person}{Zvi Reich}.}
  \bibinfo{year}{2011}\natexlab{}.
\newblock \showarticletitle{Source credibility and journalism: Between visceral
  and discretional judgment}.
\newblock \bibinfo{journal}{{\em Journalism Practice\/}} \bibinfo{volume}{5},
  \bibinfo{number}{1} (\bibinfo{year}{2011}), \bibinfo{pages}{51--67}.
\newblock


\bibitem[\protect\citeauthoryear{Rony, Hassan, and Yousuf}{Rony
  et~al\mbox{.}}{2017}]%
        {rony2017diving}
\bibfield{author}{\bibinfo{person}{Md~Main~Uddin Rony},
  \bibinfo{person}{Naeemul Hassan}, {and} \bibinfo{person}{Mohammad Yousuf}.}
  \bibinfo{year}{2017}\natexlab{}.
\newblock \showarticletitle{Diving Deep into Clickbaits: Who Use Them to What
  Extents in Which Topics with What Effects?}. In \bibinfo{booktitle}{{\em
  Proceedings of the 2017 IEEE/ACM International Conference on Advances in
  Social Networks Analysis and Mining 2017}}. ACM, \bibinfo{pages}{232--239}.
\newblock


\bibitem[\protect\citeauthoryear{Schneider}{Schneider}{2016}]%
        {schneider2016m}
\bibfield{author}{\bibinfo{person}{Schneider}.}
  \bibinfo{year}{2016}\natexlab{}.
\newblock \bibinfo{title}{Most Watched Television Networks: Ranking 2016’s
  Winners and Losers. IndieWire.}
\newblock
  \bibinfo{howpublished}{\url{http://www.indiewire.com/2016/12/cnn-fox-news-msnbc-nbc-ratings-2016-winners-losers-1201762864/}}.
    (\bibinfo{year}{2016}).
\newblock


\bibitem[\protect\citeauthoryear{Shoemaker and Reese}{Shoemaker and
  Reese}{2013}]%
        {shoemaker2013mediating}
\bibfield{author}{\bibinfo{person}{Pamela~J Shoemaker} {and}
  \bibinfo{person}{Stephen~D Reese}.} \bibinfo{year}{2013}\natexlab{}.
\newblock \bibinfo{booktitle}{{\em Mediating the message in the 21st century: A
  media sociology perspective}}.
\newblock \bibinfo{publisher}{Routledge}.
\newblock


\bibitem[\protect\citeauthoryear{Thurman}{Thurman}{2008}]%
        {thurman2008forums}
\bibfield{author}{\bibinfo{person}{Neil Thurman}.}
  \bibinfo{year}{2008}\natexlab{}.
\newblock \showarticletitle{Forums for citizen journalists? Adoption of user
  generated content initiatives by online news media}.
\newblock \bibinfo{journal}{{\em New media \& society\/}} \bibinfo{volume}{10},
  \bibinfo{number}{1} (\bibinfo{year}{2008}), \bibinfo{pages}{139--157}.
\newblock


\bibitem[\protect\citeauthoryear{Times}{Times}{2018}]%
        {nytimesdata}
\bibfield{author}{\bibinfo{person}{The New~York Times}.}
  \bibinfo{year}{(accessed October 29, 2018)}\natexlab{}.
\newblock \bibinfo{booktitle}{{\em The New York Times Annotated Corpus}}.
\newblock
\showURL{%
\url{https://catalog.ldc.upenn.edu/ldc2008t19}}


\bibitem[\protect\citeauthoryear{Turcotte}{Turcotte}{2017}]%
        {turcotte2017s}
\bibfield{author}{\bibinfo{person}{Jason Turcotte}.}
  \bibinfo{year}{2017}\natexlab{}.
\newblock \showarticletitle{Who’s Citing Whom? Source Selection and Elite
  Indexing in Electoral Debates}.
\newblock \bibinfo{journal}{{\em Journalism \& Mass Communication Quarterly\/}}
  \bibinfo{volume}{94}, \bibinfo{number}{1} (\bibinfo{year}{2017}),
  \bibinfo{pages}{238--258}.
\newblock


\bibitem[\protect\citeauthoryear{Van~Leuven and Deprez}{Van~Leuven and
  Deprez}{2017}]%
        {van2017follow}
\bibfield{author}{\bibinfo{person}{Sarah Van~Leuven} {and}
  \bibinfo{person}{Annelore Deprez}.} \bibinfo{year}{2017}\natexlab{}.
\newblock \showarticletitle{‘To follow or not to follow?’: How Belgian
  health journalists use Twitter to monitor potential sources}.
\newblock \bibinfo{journal}{{\em Journal of Applied Journalism \& Media
  Studies\/}} \bibinfo{volume}{6}, \bibinfo{number}{3} (\bibinfo{year}{2017}),
  \bibinfo{pages}{545--566}.
\newblock


\bibitem[\protect\citeauthoryear{Wikipedia}{Wikipedia}{2018}]%
        {wikifakelist}
\bibfield{author}{\bibinfo{person}{Wikipedia}.} \bibinfo{year}{(accessed
  September 24, 2018)}\natexlab{}.
\newblock \bibinfo{booktitle}{{\em List of fake news websites}}.
\newblock
\showURL{%
\url{https://en.wikipedia.org/wiki/List_of_fake_news_websites}}


\bibitem[\protect\citeauthoryear{Yamamoto, Nah, and Chung}{Yamamoto
  et~al\mbox{.}}{2017}]%
        {yamamoto2017us}
\bibfield{author}{\bibinfo{person}{Masahiro Yamamoto},
  \bibinfo{person}{Seungahn Nah}, {and} \bibinfo{person}{Deborah Chung}.}
  \bibinfo{year}{2017}\natexlab{}.
\newblock \showarticletitle{US Newspaper Editors’ Ratings of Social Media as
  Influential News Sources}.
\newblock \bibinfo{journal}{{\em International Journal of Communication\/}}
  \bibinfo{volume}{11} (\bibinfo{year}{2017}), \bibinfo{pages}{17}.
\newblock


\bibitem[\protect\citeauthoryear{Zimdars}{Zimdars}{2016}]%
        {melissa2016m}
\bibfield{author}{\bibinfo{person}{Melissa Zimdars}.}
  \bibinfo{year}{2016}\natexlab{}.
\newblock \bibinfo{title}{False, Misleading, Clickbait-y, and/or Satirical
  “News” Sources.}
\newblock
  \bibinfo{howpublished}{\url{https://docs.google.com/document/d/10eA5-mCZLSS4MQY5QGb5ewC3VAL6pLkT53V_81ZyitM/preview}}.
    (\bibinfo{year}{2016}).
\newblock


\end{thebibliography}

\end{document}